\documentclass[preprint2]{aastex}

\usepackage{epsf}

\begin{document}

\title{Cosmic-Ray Rejection by Laplacian Edge Detection}

\author{Pieter G. van Dokkum\footnote{Hubble Fellow}}
\affil{California Institute of Technology, MS~105-24, Pasadena, CA 91125}
\email{pgd@astro.caltech.edu\vspace{0.3cm}\\
{\rm \em Accepted for publication in the PASP}\vspace{-0.8cm}}

\begin{abstract}
Conventional algorithms for rejecting cosmic-rays in single
CCD exposures rely on the contrast between
cosmic-rays and their surroundings, and may
produce erroneous results if the
Point Spread Function (PSF) is smaller than the largest cosmic-rays.
This paper describes a robust algorithm for cosmic-ray rejection,
based on a variation of Laplacian edge detection.
The algorithm identifies cosmic-rays of arbitrary shapes and
sizes by the sharpness of their edges, and
reliably discriminates between poorly sampled point sources and
cosmic-rays.
Examples of its performance
are given for spectroscopic
and imaging data, including
{\em Hubble Space Telescope} WFPC2 images.
\end{abstract}


\section{Introduction}

Various methods are in use for identifying and replacing
cosmic-ray hits in CCD data. The most straightforward
approach is to
obtain multiple exposures of the same field
(or multiple non-destructive readouts during a single exposure;
e.g., Fixsen et al.\ 2000).
In general, a given
pixel will suffer from a cosmic-ray hit in only
one or two of the exposures, and remaining exposures can be used to
obtain its replacement value
(e.g., Zhang 1995). Methods for combining multiple
exposures have reached a high degree of sophistication, particularly
those developed
for dithered {\em Hubble Space Telescope} (HST) data (e.g., Windhorst,
Franklin, \& Neuschaefer 1994, Freudling 1995, Fruchter \& Hook 1997).

However, there are circumstances when cosmic-ray identification in
single exposures is required or desirable. The object of interest may
be varying or moving on short timescales, and in the case of long-slit
spectra the positions and intensities
of sky lines and object spectra may change
(e.g., Croke 1995).
Furthermore, pixels can be hit by cosmic-rays
in more than one exposure, and some affected pixels may remain
after combining individual images.
Cosmic-ray removal in individual exposures may also be desirable
if the images are shifted with respect to each other by a non-integer
number of pixels, or if the seeing varied significantly between the
exposures (see Rhoads 2000).
Finally, multiple exposures are simply not always available.

Methods for identifying cosmic-rays in single images or spectra
include median filtering (e.g., {\sc Qzap} by
M.\ Dickinson), filtering by adapted
Point Spread Functions (PSFs) (e.g., Rhoads 2000),
trained neural networks (Salzberg et al.\ 1995), and
interpolation of neighbouring pixels (e.g., the {\sc Cosmicrays}
task in the IRAF package).
All these methods effectively remove small cosmic-rays
from well sampled data.

The most widely used methods are based on some form of
median filtering, and usually include adaptations to exclude
stars and emission lines from the list of cosmic-rays.
However, problems arise when cosmic-rays affect more than half
the area of the filter, or 
the PSF is smaller than the filter.
The size of the filter is therefore a trade-off between
detecting large cosmic-rays and limiting contamination by
structure on the scale of the PSF. 

In this paper, a new algorithm for rejecting cosmic-rays in
single exposures is described. It is based on
Laplacian edge detection, which is a widely used method
for highlighting boundaries in digital images (see, e.g.,
Gonzalez \& Woods 1992). The strength of the method is that
it relies on the sharpness of the {\em edges} of cosmic-rays rather
than the contrast between entire cosmic-rays and
their surroundings. Therefore, it is largely
independent of the morphology of cosmic-rays. This property
is very useful, and forms the basis for a robust
discrimination between poorly sampled
point sources and cosmic-rays.

\section{The Laplacian}

The Laplacian of a 2-D function is a second-order derivative
defined as
\begin{equation}
\nabla^2 f = \frac{\partial f}{\partial x^2} + \frac{\partial f}
{\partial y^2}.
\end{equation}
The Laplacian is commonly used for edge detection in digital
images. In this application
the image is convolved with the Laplacian of a 2-D
Gaussian function of the form
\begin{equation}
f(x,y) = \exp \left( - \frac{r^2}{2 \sigma^2} \right),
\end{equation}
where $r^2 \equiv x^2+y^2$
and $\sigma$ is the standard deviation. The second-order
derivative with respect to $r$ has the form
\begin{equation}
\label{laplace.eq}
\nabla^2 f = \left( \frac{r^2 - 2 \sigma^2}{\sigma^4}\right)
\exp\left( - \frac{r^2}{2 \sigma^2}\right).
\end{equation}
The Laplacian has zero-crossings at $r = \pm \sqrt{2}\sigma$,
and the locations of edges are found by identifying zero-crossings
in the convolved image. The standard deviation can be tuned to
the smoothness of the edge, and by using a range
of values for $\sigma$ both sharp and smooth edges can be identified
(Marr \& Hildreth 1980).

Cosmic-rays have very sharp edges, and the convolution kernel
should be most sensitive to variations on small scales.
The appropriate discrete
implementation of Eq.\ \ref{laplace.eq} has the form
\begin{equation}
\label{kernel.eq}
\nabla^2 f = \frac{1}{4} \left\{ \begin{array}{rrr} 0 & -1 & 0 \\
-1 & 4 & -1 \\ 0 & -1 & 0 \end{array} \right\}.
\end{equation}
The average value of a Laplacian image (obtained by convolving
an image with the kernel given in Eq.\ \ref{kernel.eq}) 
is zero, and smooth structure in the image is removed.

\section{Implementation}

\subsection{Basic Procedure}

A straightforward convolution of Eq.\ \ref{kernel.eq} with a CCD image
produces negative cross patterns around high pixels. As a result,
connected cosmic-ray pixels
suffer from attenuation by the negative cross patterns of their
neighbors.  Hence before convolution the original
image $I$ needs to be
subsampled. The results
are independent of the subsampling factor, and a factor two
is computationally least expensive:
\begin{equation}
I_{i,j}^{(2)} = I_{{\rm int}([i+1]/2),{\rm int}([j+1]/2)},
\end{equation}
with $n \times n$ the size of the original image
and $i,j = 1, \ldots, 2n$. 
The Laplacian of the subsampled image is
\begin{equation}
{\cal L}^{(2)} = \nabla^2 f \circ I^{(2)} ,
\end{equation}
with $\circ$ denoting convolution.

The Laplacian of the edge of a cosmic-ray is negative on the outside
and positive on the inside of the cosmic-ray.  Therefore, by setting
all negative values in the Laplacian image to zero
cosmic-ray affected pixels are
retained and their negative cross patterns are removed:
\begin{equation}
{\cal L}^{(2)+} = \left\{ \begin{array}{ll}
{\cal L}^{(2)} & \hbox{if} \quad {\cal L}^{(2)} \geq 0
\\
0 & \hbox{if} \quad {\cal L}^{(2)} < 0
\end{array} \right.
\end{equation}
Finally, the image is resampled to its original resolution:
\begin{equation}
{\cal L}^+_{i,j} = \frac{1}{4}
\left( {\cal L}^{(2)+}_{2i-1,2j-1} +
{\cal L}^{(2)+}_{2i-1,2j} +
{\cal L}^{(2)+}_{2i,2j-1} +
{\cal L}^{(2)+}_{2i,2j} \right),
\end{equation}
with  $i,j = 1, \ldots, n$. 

\begin{figure}[htb]
\epsfxsize=7.6cm
\epsfbox{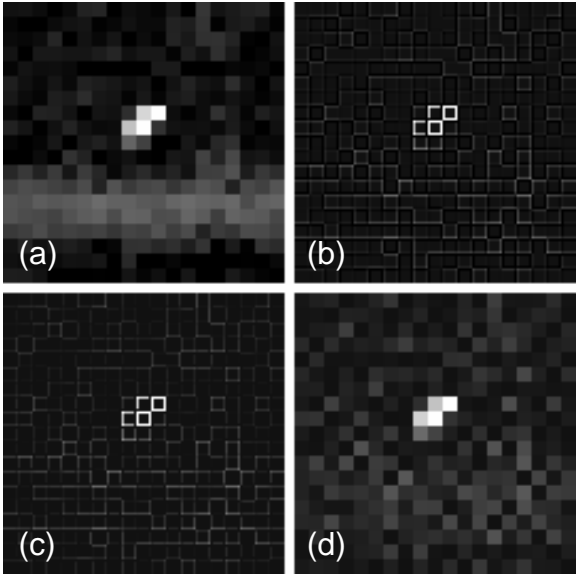}
\caption{\small
Illustration of Laplacian edge detection.
The original image is shown in (a). Panel (b) shows the same
image after subsampling by a factor six and convolution with
the Laplacian kernel. Edges are positive on the inside
of the cosmic-ray, and negative on the outside.
Negative pixels are set to zero in (c), and the image is
block averaged to its original resolution in (d).
\label{explain.plot}
}
\end{figure}

The process is
illustrated in Fig.\ \ref{explain.plot}. A small section
of a 2D long slit spectrum is shown
in (a). Panel (b) shows the same image, subsampled by a factor six
and convolved with the Laplacian kernel given in Eq.\ \ref{laplace.eq}.
Edges are at the locations of zero-crossings. In panel (c)
all negative values are set to zero. Finally, the image is
block averaged by a factor six (panel d). The cosmic-ray stands
out clearly. Because the spectrum is smooth on scales of $\sim 1$
pixel its Laplacian is close to zero.

The numerical value associated with each edge is
the difference between the two neighbouring pixels, and
the resampled Laplacian of an image $I$ consisting of a
smooth background $B$ with superposed noise and
cosmic-rays is approximately
\begin{equation}
{\cal L}^+ \sim \left\{ \begin{array}{ll}
f_s(I - B) & \hbox{if} \quad I-B\geq 0 \\
0 & \hbox{if} \quad I-B<0
\end{array} \right.
\end{equation}
with $f_s$ the subsampling factor that was used.
The Laplacian thus retains the
flux in high pixels and removes the local background, making it
very useful for identifying cosmic-rays.

\begin{figure}[htb]
\epsfxsize=7.6cm
\epsfbox{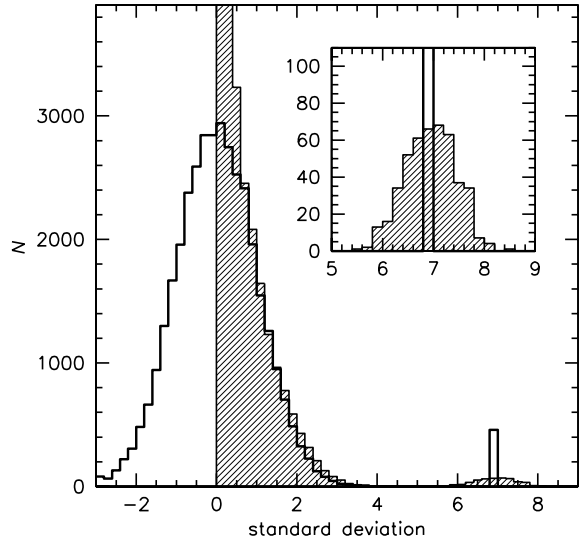}
\caption{
\small
The thick solid line shows the noise properties of a simulated
image with $\sim 10^3$ $7 \sigma$
cosmic-rays. The hatched histogram shows the
distribution of pixel values in the Laplacian image.
Pixel values
in the Laplacian image were divided by the subsampling factor.
In the positive tail of the noise distribution
the characteristics of the original image are (to
a good approximation) conserved.
The width of the distribution of cosmic-rays
is increased by $0.5 \sigma$.
\label{his.plot}
}
\end{figure}

To identify cosmic-rays in the Laplacian
image the value of each pixel is compared to the expected noise
at that location.
The noise characteristics of the Laplacian image are not the same
as in the original image: the Laplacian operator itself
increases the noise by a factor $\sqrt{5/4}$, and
all negative Poisson fluctuations in the
original image are (close to) zero in the Laplacian image.
These effects are
demonstrated in Fig.\ \ref{his.plot} for
a simulated image containing $\sim 10^3$
$7 \sigma$ cosmic-rays.
As expected, the distribution of
pixel values near zero is strongly distorted by
the Laplacian. However, the
positive tail of the distribution (i.e., the
region where cosmic-rays are found)
remains virtually unchanged. Therefore
the noise properties of the original image can be used
for identifying cosmic-rays in ${\cal L}^+$, which greatly simplifies
the analysis.

The noise model is constructed by convolving
the original image with a median filter:
\begin{equation}
\label{noise.eq}
N = g^{-1} \sqrt{g \left( M_5 \circ I \right) + \sigma_{rn}^2},
\end{equation}
where $g$ is the gain in electrons/ADU, $\sigma_{rn}$ is the
readnoise in electrons, and $M_5$ is a $5 \times 5$ median filter.
For long slit spectroscopic data the algorithm offers the option of
fitting and subtracting sky lines and/or object spectra. These fits
provide the basis for the long slit noise model, which is
optimized by applying Eq.\ \ref{noise.eq}
to the residuals from the fits.

The Laplacian image is divided by the noise model to obtain the
deviations from the expected Poisson fluctuations:
\begin{equation}
\label{sigmap.eq}
S = \frac{{\cal L}^+}{f_s N}.
\end{equation}
Cosmic-rays are identified by selecting
pixels in $S$ that are above a given threshold $\sigma_{\rm lim}$.

For single-pixel cosmic-rays on a smooth background
the detection probability depends on
the noise in the background $\sigma_{\rm org}$ only.
The error in the background
estimate scales as $\sigma_{\rm org}/\sqrt{n}$, with $n$ the
number of pixels. Since four neighbouring
pixels are used to determine the Laplacian of a given pixel,
the width of the distribution of cosmic-rays is increased by
$\sigma_{\rm org}/2$ (see Fig.\ \ref{his.plot}).
As an example, if a detection threshold of $5 \sigma$
is applied, $\sim 5$\,\% of $4 \sigma$ peaks in the original image
will be marked as cosmic-rays,
$\sim 50$\,\% of $5 \sigma$ peaks, and
$\sim 95$\,\% of $6 \sigma$ peaks.

The detection probability of cosmic-rays larger than a single pixel
depends on the number of connected pixels and the pixel-to-pixel
variation within the cosmic-ray. In general the Laplacian
(and hence the detection probability) is lowest for cosmic-rays with
small pixel-to-pixel variations. In the limiting case of
a cosmic-ray with negligible pixel-to-pixel
variation
\begin{equation}
S_{i,j} \sim N_{i,j}^{-1}
\left( 1- \frac{n_{i,j}}{4}\right) (I_{i,j}-B_{i,j}),
\end{equation}
with $n_{i,j}$ the number of cosmic-ray pixels adjacent to
pixel $(i,j)$.
Pixels on the corners of large cosmic-rays have at most two
adjacent pixels, and the Laplacian
always retains at least $\approx 50$\,\% of their flux.
Therefore, (arbitrarily)
large cosmic-rays
can be removed by applying the rejection process
iteratively. Before each iteration, pixels flagged as cosmic-rays
in previous iterations
are replaced by the median of surrounding ``good'' pixels.acd

\subsection{Removal of Sampling Flux}
\label{sampling.sec}

The Laplacian gives the difference between one pixel and its neighbours,
and contains no information on the nature of detected features.
Real astronomical objects produce signal in the Laplacian image
because of Poisson noise and because their intrinsically
smooth intensity profiles are sampled by the pixels. This ``sampling flux''
is generally small, and will not produce spurious detections
as long as it does not exceed the predicted
Poisson fluctuations (see Eq.\ \ref{noise.eq},
\ref{sigmap.eq}). However, for bright objects the sampling flux
can be significant, in particular if the Point Spread
Function (PSF) is not well sampled.

Sampling flux is removed from $S$ in two steps.
First, all structure that is smooth on scales of $\gtrsim 5$
pixels is removed by a $5 \times 5$ median filter:
\begin{equation}
S' = S - (S \circ M_5).
\end{equation}
This procedure very effectively removes sampling flux resulting from
extended bright objects (including point sources if the PSF is
well sampled). Because of the large size of the filter
cosmic-rays and the noise properties of $S$ remain unaffected.

Next, sampling flux resulting from critically sampled
(or even undersampled) point sources is removed.
As is well known, it is very hard to
distinguish cosmic-rays from stars and emission lines in
marginally sampled data, because they
can have very
similar pixel-to-pixel variations within an area $\lesssim 3
\times 3$ pixels.

Point sources are distinguished from cosmic-rays by
their symmetry. An image only containing
symmetric fine structure on scales of $2 - 3$ pixels
is constructed and compared to the Laplacian image.
This ``fine structure image'' $\cal F$ is
created from the original image by a combination of median filters:
\begin{equation}
{\cal F} = (M_3 \circ I) - \left( \left[ M_3 \circ I \right] \circ M_7 \right),
\end{equation}
where $M_n$ is an $n \times n$ median filter. The second term serves
to remove large scale structure from $\cal F$.
An important property of $\cal F$ is that central pixels of
undersampled point sources do not vanish, but
are replaced by the median of the surrounding pixels.
The Laplacian image is divided by
the fine structure image, and cosmic-rays are selected as
those pixels which have $S' > \sigma_{\rm lim}$ {\em and}
${\cal L}^+ / {\cal F} > f_{\rm lim}$, with $f_{\rm lim}$
defining the minimum contrast between the Laplacian image and
the fine structure image.

\begin{figure}[htb]
\epsfxsize=7.6cm
\epsfbox{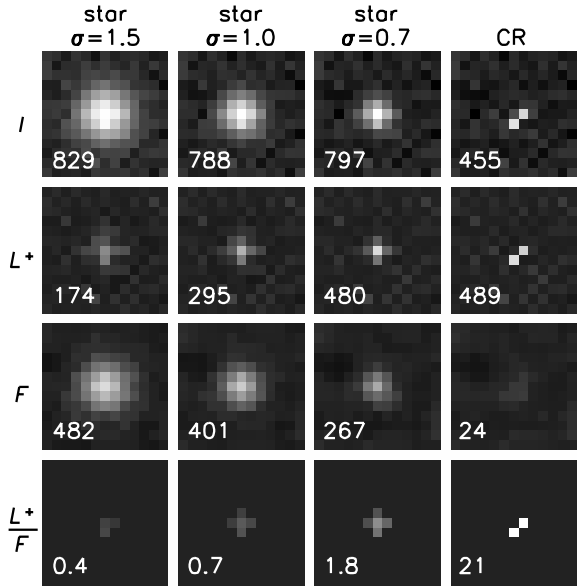}
\caption{\small
Differentiating between marginally sampled point
sources and cosmic-rays.
Panels show, from top to bottom, artifical images of stars and
a cosmic-ray, the Laplacian of these images ${\cal L}^+$, their
fine structure image $\cal F$, and the Laplacian divided
by the fine structure ${\cal L}^+ / {\cal F}$. The number
in each panel is the value of the highest pixel. The highest
pixels in
the Laplacian images of the undersampled star
($\sigma = 0.7$ pixels) and the cosmic-ray
are similar. However, they
are very different
after division by the fine structure image.
\label{stars.plot}
}
\end{figure}

Figure\ \ref{stars.plot} demonstrates the procedure.
Artificial images of three stars and a cosmic-ray are shown in
the top row.
One stellar image is slightly oversampled ($\sigma = 1.5$ pixels),
the second is critically sampled ($\sigma = 1.0$ pixels), and
the third is slightly undersampled ($\sigma = 0.7$ pixels).
The contrast between the cosmic-ray and its local background is
identical to the contrast between the central pixel of the
undersampled star and its local background.
As a result, the highest pixels in
their Laplacian images ${\cal L}^+$ (shown
in the second row) have very similar values.
The third row shows fine structure images $\cal F$.
The cosmic-ray has very low signal in $\cal F$, whereas 
the stars retain a significant fraction of their flux because
of their symmetry. The bottom row shows the Laplacian images
divided by the fine structure, ${\cal L}^+ / \cal F$.
Only pixels with ${\cal L}^+ / {\cal F} > f_{\rm lim}$
are retained.
The critically sampled star has ${\cal L}^+ / {\cal F}
= 0.7$, and ${\cal L}^+ / {\cal F}=1.8$ for the
undersampled star. The cosmic-ray has ${\cal L}^+ / {\cal F}=21$, and
is easily distinguished from undersampled point sources.

\begin{figure}[htb]
\epsfxsize=7.6cm
\epsfbox{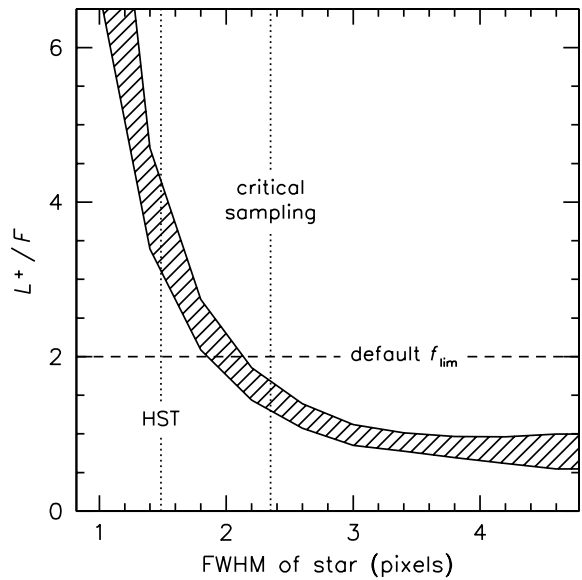}
\caption{\small
Dependence of the
ratio of the Laplacian ${\cal L}^+$
and the fine structure image $\cal F$ on the sampling,
as derived from simulations (see text).
For a given FWHM of stars the value of
$f_{\rm lim}$ should be chosen such that it exceeds the dashed
band. The default $f_{\rm lim} = 2$ is appropriate for data that
are critically, or better, sampled. HST WFPC2
data require $f_{\rm lim} \approx 5$.
\label{sim.plot}
}
\end{figure}

In general, the appropriate value of $f_{\rm lim}$
depends on the sampling of a given point source,
its S/N ratio, and whether it
lies on the center of a pixel or close to the edge.
These effects can be simulated by creating artificial PSFs
with varying sampling, S/N ratios, and subpixel positions, and
calculating ${\cal L}^+ / {\cal F}$ for each star. Figure \ref{sim.plot}
shows the dependence of
${\cal L}^+ / {\cal F}$ on the sampling, expressed as
the full width at half maximum (FWHM) of stars (in pixels). The width
of the shaded region demonstrates the (maximum)
effects of varying S/N ratio and subpixel position.
These simulations
demonstrate that the
value of ${\cal L}^+ / {\cal F}$, and hence the appropriate choice
of $f_{\rm lim}$, depends mainly on the sampling. For data that are
well sampled $f_{\rm lim} = 2$ is appropriate; hence this is
the default value in the algorithm.
For undersampled data higher values of $f_{\rm lim}$ are needed
to discriminate point sources and cosmic-rays;
as an example, data taken with
the Wide Field chips in HST's WFPC2 camera require
$f_{\rm lim} \approx 5$.

\subsection{Additional Features}

The basic algorithm detects cosmic-rays and rejects point sources
from the initial list. In addition,
the program {\sc L.A.Cosmic} allows a lower detection
threshold to be used for pixels neighbouring those already flagged as
cosmic-rays. It also replaces cosmic-rays by the median
of surrounding ``good'' pixels, and offers the option of
applying the algorithm iteratively. On a Sun UltraSparc 1 (200\,MHz)
the IRAF implementation of {\sc L.A.Cosmic}
requires approximately 65\,s per iteration for an image of
$800 \times 800$ pixels. The run time scales linearly
with the number of pixels.

\section{Examples}
The algorithm was tested on a variety of real and artificial data sets,
consistently producing very good results.
The examples given here serve to illustrate its performance.

\subsection{Well Sampled Imaging CCD Data}
\begin{figure}[htb]
\epsfxsize=16.5cm
\epsfbox{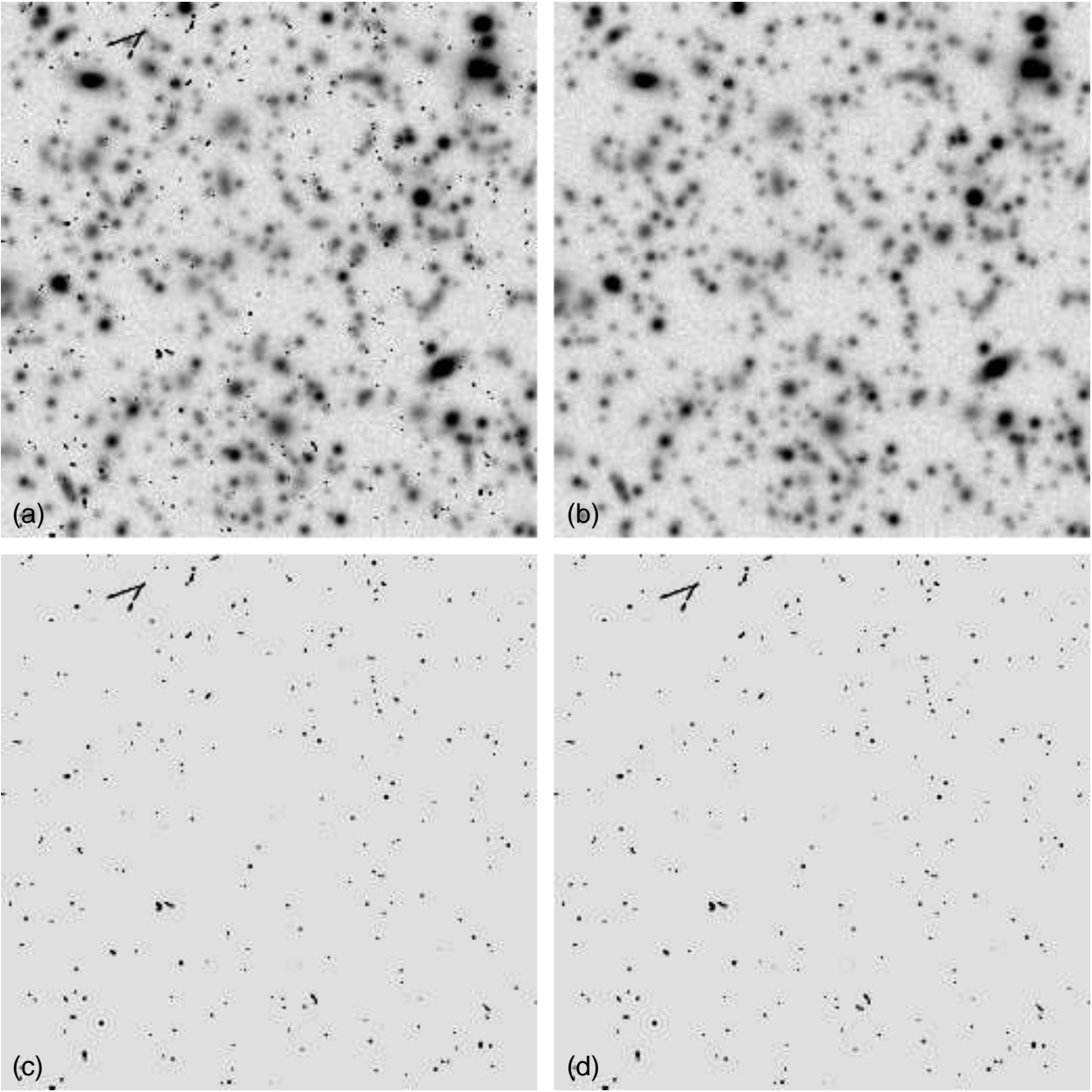}
\caption{\small
(a) Artificial 
image containing 500 stars, 100 galaxies,
and 227 cosmic-rays. (b) Reconstruction of the image by {\sc L.A.Cosmic}.
The true cosmic-ray image is shown in (c), and the cosmic-rays found
by {\sc L.A.Cosmic} are shown in (d).
\label{example1.plot}
}
\end{figure}

Figure \ref{example1.plot}(a) shows a well sampled 
artifical image containing 500 stars, 100 galaxies, and 227 cosmic-rays.
Stars have $\sigma =
1.5$ pixels, equivalent to, for example,
FWHM\,$= 0\farcs 78$ seeing with $0\farcs 22$ pixels.
All cosmic-rays are $\geq 5 \sigma$ above the sky background.
The reconstruction by {\sc L.A.Cosmic} is shown in (b).
Panel (c) shows the input cosmic-ray image, and panel (d)
shows the cosmic-rays found by {\sc L.A.Cosmic}.
The program found 222 of the 227 cosmic-rays (98\,\%).
Importantly, only 1 of the 500 stars (0.2\,\%) and none of the galaxies
was inadvertently identified as a cosmic-ray.

For well sampled imaging data
the performance is similar to median filtering methods such
as {\sc Qzap} (by M.\ Dickinson). Sophisticated median filtering
recovers close to 100\,\% of cosmic-rays that are smaller than
the filter size, but breaks down if the cosmic-ray is larger
than the filter, and/or the FWHM of the
PSF is smaller than the filter.

\subsection{HST WFPC2 Data}

Cosmic-rays in
images obtained with WFPC2 on HST
are notoriously difficult
to remove, because of their large number and the
undersampling of the PSF. Nevertheless, the algorithm described here performs
very well on WFPC2 data. The method is insensitive to the size of
cosmic-rays, and the undersampling of the PSF can be taken into
account by setting the parameter $f_{\rm lim} = 5$
(see Sect.\ \ref{sampling.sec}).

\begin{figure}[htb]
\epsfxsize=16.5cm
\epsfbox{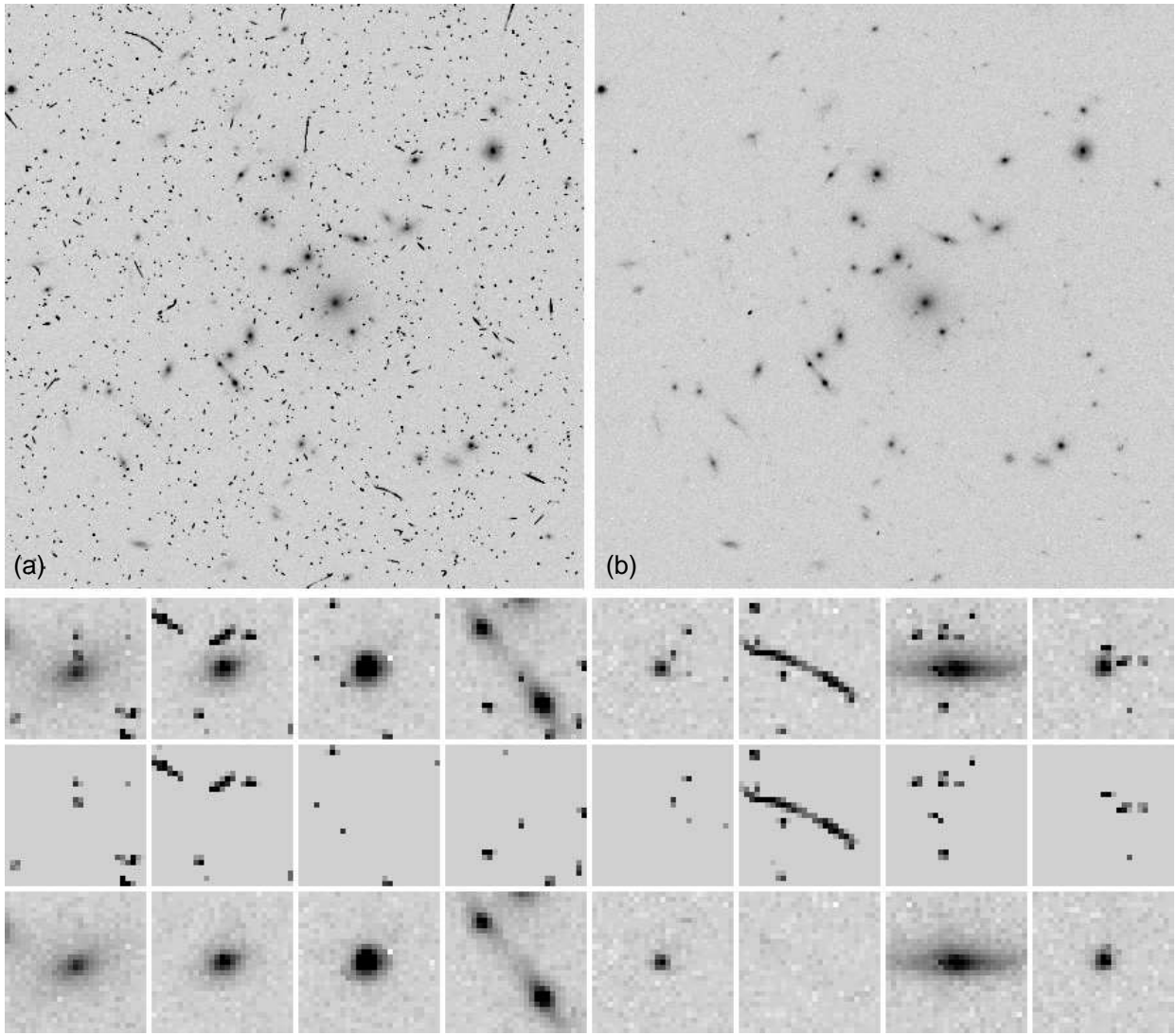}
\caption{\small
(a) HST WFPC2 image of galaxy cluster MS\,1137+67. The
restoration by {\sc L.A.Cosmic} is shown in (b). Small panels
show close-ups for a selection of stars and galaxies in various WFPC2
images. The algorithm leaves stars intact, and is able to remove
cosmic-rays of arbitrary shapes and sizes.
\label{example2.plot}
}
\end{figure}

Figure \ref{example2.plot}(a) shows part of
a 2400\,s WF observation in $I_{F814W}$ of galaxy cluster MS\,1137+67.
The reconstruction of the image by
{\sc L.A.Cosmic} is shown in (b).
\vspace{1.6cm}\\

Virtually all
cosmic-rays are removed, and none of the real objects is mistaken
for a cosmic-ray. The small panels show examples of stars and galaxies
in WF chips, extracted from WFPC2 observations of various targets.
The algorithm leaves stars intact, and is able to remove
arbitrarily large cosmic-rays.

The reliability of cosmic-ray identification can be tested by comparing
the results of {\sc L.A.Cosmic} on single images to ``true''
cosmic-ray images created from multiple exposures. The test used
one of the CR--SPLIT WFPC2
images of the cluster MS\,2053--04 ($z=0.58$). As a result of
its low Galactic latitude approximately 50\,\%
of $I_{\rm F814W}<22$ objects are stars. The
reduction of these data is described in van Dokkum et al.\ (2001).
The algorithm found 5638 (98.1\,\%) of 5750 cosmic-ray affected pixels
deviating more than $6 \sigma$ from the background,
and 4687 (99.1\,\%) of 4729 pixels deviating more than $10 \sigma$.
The number of false positives, i.e., pixels inadvertently marked
as cosmic-rays, is 72 (1.2\,\%) at $\geq 6 \sigma$, and only 1
(0.02\,\%) at $\geq 10 \sigma$. These numbers compare favorably to
cosmic-ray rejection algorithms based on morphological classification by
neural networks (Salzberg et al.\ 1995).

\subsection{Spectroscopic CCD Data}

The algorithm optimized for
spectroscopic long-slit data is very similar to the implementation
for imaging data. The main difference is that the program
offers the possibility of fitting and subtracting sky lines
and the object spectrum before convolution with the Laplacian kernel.

\begin{figure}[htb]
\epsfxsize=16.5cm
\epsfbox{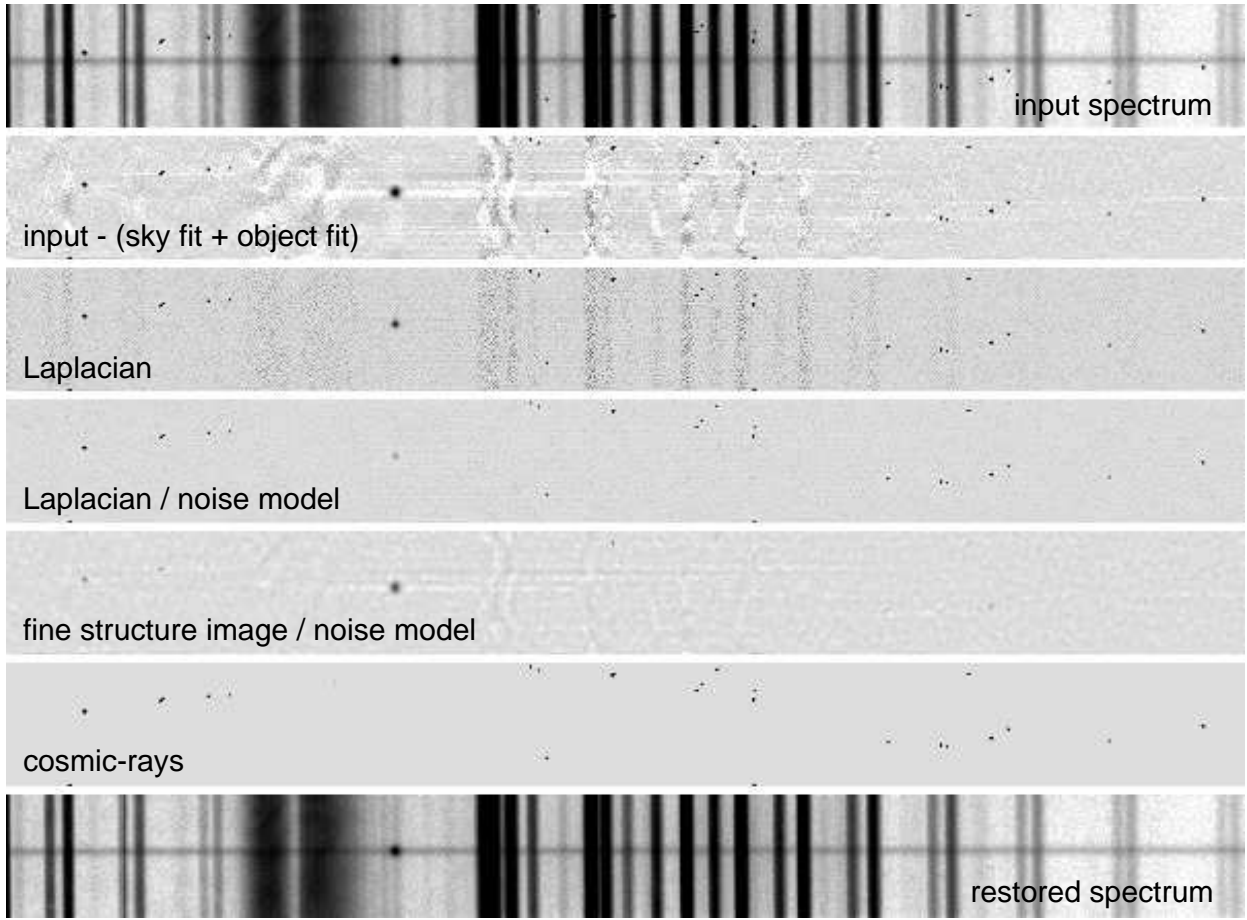}
\caption{\small
Demonstration of Laplacian cosmic-ray rejection for long-slit
spectra. From top to bottom are shown: the original spectrum,
the residuals after subtraction of 1D-fits to the sky lines and the
object spectrum, the Laplacian image, the Laplacian image divided
by the noise model, the finestructure image, the bad pixel map,
and the restored spectrum. Note that the Laplacian very effectively removes
fringing. In this case all cosmic-rays are removed in a single
iteration. The bright emission line is not marked
as a cosmic-ray because of its prominence in the fine structure image.
\label{example3.plot}
}
\end{figure}

An example (an 1800\,s long slit spectrum of a galaxy,
obtained with the Low Resolution
Imaging Spectrograph on the W. M. Keck Telescope)
is shown in Fig.\ \ref{example3.plot}.
Note that the Laplacian is very effective in removing the fringe pattern
that is present after subtracting strong sky lines. The fine structure
image $\cal F$ is used to identify emission lines and other sharp
features in the spectra, in similar fashion as the identification
of undersampled stars in imaging data.

\onecolumn

\twocolumn
\section{Conclusions}

Cosmic-rays in single images or spectra can be removed by a variation
of Laplacian edge detection. The procedure is robust, and requires
very few user-defined parameters. The method rejects
cosmic-rays of arbitrary size
and distinguishes undersampled
point sources from cosmic-rays with high confidence.
It is implemented in the program
{\sc L.A.Cosmic}, which can be obtained from
http://www.astro.caltech.edu/\~{}pgd/lacosmic/ .



\acknowledgments

I thank Martin Zwaan for his contributions in the initial phase of
this study, and Josh Bloom for discussion and comments. 
The insightful comments and suggestions of
the referee, James Rhoads, improved the paper.
The author acknowledges support by
NASA through Hubble Fellowship grant HF-01126.01-99A awarded by the
Space Telescope Science Institute, which is operated by the
Association of Universities for Research in Astronomy, Inc., for NASA
under contract NAS 5-26555.

\newpage

\end{document}